# HARD DIFFRACTION AT HERA AND THE GLUONIC CONTENT OF THE POMERON

A. Capella, A. Kaidalov*, C. Merino, D. Pertermann** and J. Tran Thanh Van

Laboratoire de Physique Théorique et Hautes Energies***

Bâtiment 211, Université de Paris-Sud, 91405 Orsay cedex, France

**Abstract**

We show that the previously introduced CKMT model, based on conventional Regge theory, gives a good description of the HERA data on the structure function $F_2^D$ for large rapidity gap (diffractive) events. These data allow, not only to determine the valence and sea quark content of the Pomeron, but also, through their $Q^2$ dependence, give information on its gluonic content. Using DGLAP evolution, we find that the gluon distribution in the Pomeron is very hard and the gluons carry more momentum than the quarks. This indicates that the Pomeron, unlike ordinary hadrons, is a mostly gluonic object. With our definition of the Pomeron flux factor the total momentum carried by quarks and gluons turns out to be $0.3 \div 0.4$ - strongly violating the momentum sum rule.

\* Permanent address : ITEP, B. Cheremushkinskaya ulitsa 25, 117259 Moscow, Russia
\*\* Permanent address : University of Siegen, 5900 Siegen, Germany
\*\*\* Laboratoire associé au Centre National de la Recherche Scientifique. URA D0063

## 1. Introduction

Diffraction dissociation of virtual photons, observed by the H1 and ZEUS collaborations at HERA [1-4] provides information on the nature of the Pomeron and on its partonic structure. In a previous paper [5] a model (CKMT model) based on Regge theory has been proposed for the description of diffractive dissociation of both real and virtual photons. In this model the structure function of the Pomeron $F_P(\beta, Q^2)$ is related to that of the deuteron [6] via Regge factorization. Recently, experimental information on the Pomeron structure function at different values of $Q^2$ has been obtained [2, 4]. These data are in reasonable agreement with the predictions of the model.

In this paper, we extend our analysis of the Pomeron structure function and determine the gluonic content of the Pomeron. We use our model as initial condition in DGLAP evolution. We show that the $Q^2$-dependence of the diffractive structure function, $F_2^D$, observed experimentally, can only be understood if the gluon distribution in the Pomeron is very hard.It is also shown that the gluons carry more momentum than quarks. This observation confirms theoretical expectations that the Pomeron is mainly connected with gluonic degrees of freedom.

The plan of the paper is as follows. In Section 2 we recall the ingredients of the CKMT model. In Section 3 we perform the QCD evolution and in Section 4 we present our numerical results. In Section 5 we discuss other recent QCD analysis and give our conclusions.

## 2. - The CKMT model

The contribution of diffractive processes to the deep inelastic scattering (DIS) cross-section, corresponding to the Pomeron exchange diagram of Figs. 1 and 2, can be written in the form [4]

$$\frac{d^4\sigma^D}{dx\, dQ^2\, dx_P\, dt} = \frac{4\pi\alpha_{em}^2}{x\, Q^4}\left\{1 - y + \frac{y^2}{2\left[1 + R^D(x, Q^2, x_P, t)\right]}\right\} F_2^D(x, Q^2, x_P, t) \quad (1)$$

where $x$ and $y$ are standard DIS variables, $t$ is the invariant momentum transfer $t =$

$-(p-p')^2$ and $x_P$ is the fraction of the proton momentum carried by the Pomeron :

$$x_P = \frac{p_P \cdot p_\gamma}{p_p \cdot p_\gamma} \simeq \frac{M^2 + Q^2}{W^2 + Q^2} \equiv \frac{x}{\beta} \quad . \tag{2}$$

We have argued in Ref. [5] that the Pomeron in the diagrams of Figs. 1 and 2 can be considered as a Regge pole with a trajectory $\alpha_P(t) = \alpha_P(0) + \alpha' t$ determined from an analysis of soft processes, in which absorptive correction (Regge cuts) are taken into account. We have [5], [7]

$$\alpha_P(0) = 1.13 \quad , \quad \alpha'_P(0) = 0.25 \text{ GeV}^{-2} \quad . \tag{3}$$

In this case the diffractive contribution to DIS can be written in a factorized form

$$F_2^D(x, Q^2, x_P, t) = \frac{\left(g_{pp}^P(t)\right)^2}{16\pi} x_P^{1-2\alpha_P(t)} F_P(\beta, Q^2, t) \tag{4}$$

where $g_{pp}^P(t)$ is the Pomeron-proton coupling $g_{pp}^P(t) = g_{pp}^P(0)\exp(Ct)$ with $(g_{pp}^P(0))^2 = 23$ mb and $C = 2.2$ GeV$^{-2}$. Eq. (4) is the definition* of the Pomeron structure function $F_P(\beta, Q^2, t)$ with the variable $\beta = x/x_P$ playing the same role as the Bjorken variable $x$ in $F_2^p$. As emphasized in Ref. [5], the definition of $F_P$ depends on the particular choice of the Pomeron flux factor. For instance in Ref. [8] the flux factor differs from ours by a factor $2/\pi$. (See also Ref. [9]).

At large $Q^2$ we have

$$F_P(\beta, Q^2, t) = \sum_i e_i^2 \beta \left[q_i^P(\beta, Q^2, t) + \bar{q}_i^P(\beta, Q^2, t)\right] \quad . \tag{5}$$

Due to the arbitrariness in the normalization of $F_P$ discussed above, the partonic distributions $q_i^P$ and $\bar{q}_i^P$ do not satisfy, in general, the energy-momentum sum rule.

Many models of the Pomeron structure function are available in the literature [8,10-14]. In the CKMT model [5] $F_P$ is determined using Regge factorization together with the values of the triple Regge couplings determined from soft diffraction data. More precisely, the proton and Pomeron structure functions are given in terms of the diagrams of Figs. 3 and 4, respectively. (Note that Fig. 4 is the upper part of the diagram obtained by squaring, in the sense of unitarity, the $\gamma P$ amplitude in Fig. 1). We see that, due to factorization, $F_2^P$ can be obtained from $F_2^p$ (or more precisely from the combination $F_2^d = \frac{1}{2}(F_2^p + F_2^n)$), by replacing the Reggeon-proton couplings $g_{pp}^P$ and $g_{pp}^f$ appearing in the lower part of Fig. 3, by the corresponding triple Reggeon couplings $r_{PPP}$ and $r_{PPf}$ which appear in the lower part of Fig. 4. The latter can be determined from soft diffraction data (see below). Another difference between the two structure functions is the $\beta \to 1$ behaviour. Indeed, as far as the dimensional counting rules are concerned, the Pomeron is like a pion and, therefore, there is one spectator less than in the proton case. Thus the power of $1 - \beta$ is smaller by two units in the Pomeron case. (The same result is obtained at $Q^2 = 0$ in terms of Regge intercepts). The CKMT model can be summarized with the following formulae [5, 6], valid in the region $1 \leq Q^2 \leq 5$ GeV$^2$ :

$$F_P(\beta, Q^2) = F_2^d \left(\beta, Q^2; A \to eA, B \to fB, n \to n - 2\right) \quad , \tag{6}$$

$$F_2^d(x, Q^2) = A(Q^2) \, x^{-\Delta(Q^2)} \, (1-x)^{n(Q^2)+4} + B \, x^{1-\alpha_R(0)} \, (1-x)^{n(Q^2)} \tag{7}$$

where $e = r_{PP}^P(0)/g_{pp}^P(0)$ and $f = r_{PP}^f(0)/g_{pp}^f(0)$. The values of all other parameters are given in Refs. [5] [6]. A main feature of the CKMT model is the $Q^2$ dependence of the effective Pomeron intercept, $\Delta \cong \alpha_P^{eff} - 1$, which appears in the upper part of the diagrams in Figs. 3 and 4. It was argued in Ref. [6] that this is due to the fact that the size of the absorptive corrections decreases when $Q^2$ increases. As a consequence, when using a parametrization with a simple power $x^{-\Delta}$, $\Delta$ must depend on $Q^2$.

---

* Note that the definition of $F_P$ in Eq. (4) differs from the one in Ref. [5] by a factor $(1 - \beta)$. The definition in Eq. (4), which coincides with the one used in experimental papers [2, 4], is more appropriate for a partonic interpretation. We thank X. Artru for an enlightening correspondence on this point.

The parameters $e$ and $f$ can be determined from soft diffraction data. Here also absorptive corrections, which are very important in diffractive processes [15], have to be taken into account when extracting the values of $e$ and $f$ from the data. In an analysis of soft diffraction without absorption [16] one obtains e $\cong$ f $\cong$ 0.025. In the first paper of Ref. [15] it was shown that these values have to be multiplied by a factor of about three in order to take into account absorptive effects. The values of e and f have, of course, some uncertainties-the largest being in the value of $f$ [5]. Experimentally, the $t$-dependence of the triple Reggeon couplings is very small. We have taken it to be the same for $r_{PP}^P$ and $r_{PPf}$ and included it in the function $C$. In this way the $t$-dependence of $F_P$, which is expected to be very small, is factored out.

The comparison of the prediction [5] of the CKMT model with recent data from the H1 and ZEUS collaborations is quite satisfactory. The data confirm the factorized dependence of the hard diffractive cross-section on $x_P$ (Eq. (1)) with the parameter of the Pomeron trajectory in Eq. (2). Agreement in absolute values is also good for most values of $\beta$ and $Q^2$, favoring our hypothesis that soft and hard diffraction are governed by the same triple Regge couplings.

## 3. QCD evolution

A closer look at the model predictions shows however a systematic decrease with $Q^2$ at $\beta \geq 0.2$. Such a trend is not present in the data. The reason for such a decrease is the following. As explained in Ref. [5] the model predictions at large $Q^2$ have been obtained from DGLAP evolution [17], using Eqs. (6) and (7) at $Q_0^2$ as initial condition. The result depends, of course, on the input gluon distribution function. In Ref. [5] it was assumed that the relation between gluon and sea quark distributions in the Pomeron was the same as in the proton. This corresponds to a rather soft gluon distribution in $(1-\beta)^3$ at $Q^2 \approx$ 5 GeV$^2$. It was already pointed out in Ref. [5] that this assumption was not justified and should be changed. Indeed, such a behaviour is valid for ordinary mesons with a large valence quark content. However, the Pomeron can have a much harder gluon distribution ("valence gluons"). This is also in agreement with UA8 data [18].

In the following we will perform DGLAP $Q^2$-evolution using the valence and sea quark initial distribution given by Eqs. (6) and (7) at $Q_0^2 = 5$ GeV$^2$, with a harder gluon distribution. More precisely, we use the same $\beta \to 0$ behaviour as for sea quark [5], but we leave the power $n_g$ of $1 - \beta$ as a free parameter, i.e. we put

$$\beta \, g_P^g(\beta) = e \, A_g \, \beta^{-\Delta(Q_0^2)} (1-\beta)^{n_g(Q_0^2)} \tag{8}$$

where $A_g = 1.71$ is the gluon normalization constant in a nucleon [6] and $e$ and $\Delta$ are defined in Eq. (7). Note that the gluon normalization is obtained from Regge factorization and the energy-momentum sum rule is not used. For $n_g$ small enough, the $Q^2$ behaviour of $F_P$ will be very different from the one of $F_2^p$. The latter exhibits a decrease with $Q^2$ for $x \geq 0.2$ due to a softening of the quark distribution resulting from gluon emission. Even though the same effect is present for the Pomeron, a harder distribution of gluons will lead in this case, through gluon decay, to an increase with $Q^2$ in the number of quarks up to comparatively large values of $\beta$.

Numerical calculations have been performed using the QCD evolution program of Refs. [19, 20, 14]. We present the results in a one loop approximaton but we have checked that practically the same results are obtained in two loops. The input and evolved partonic distributions are given in Figs. 5-8.

## 4. Numerical results

We have found [21] that a good description of the $Q^2$ dependence of the HERA data is achieved with either $n_g = 0$ or $-0.5$ at $Q_0^2 = 5$ GeV$^2$. We have also allowed the parameters $e$ and $f$ in Eqs. (6) and (8) to vary within limits consistent with the results from the analysis of soft diffraction. A good description of the data is obtained with $e = f = 0.07$. All other parameters in Eq. (7) have the values given in Ref. [6].

The 3-dimensional structure function $F_2^D(x, Q^2, x_P)$, integrated over $t$, is plotted versus $x_P$ in Fig. 9 and compared with H1 data. The results for the $Q^2$ and $\beta$ dependence of $F_2^D$ (integrated over $t$ and $x_P$) are given in Fig. 10 and compared with H1 data. In both cases, unpublished data from the ZEUS collaboration are also available. These data are in agreement, within errors, with the ones of H1. We see that the $Q^2$-dependence at large $\beta$ is flat, in agreement with experiment - whereas in Ref. [5] there was a substantial decrease with increasing $Q^2$.

In Fig. 11 we give the Pomeron structure function defined in Eq. 4 at different values of $Q^2$. We see that the cross-over, which is at $\beta \sim 0.2$ when the gluon distribution in the Pomeron is taken to be proportional to that of the proton [5], is shifted to larger values of $\beta$. Comparison of Figs. 11a ($n_g = 0$) and 11b ($n_g = -0.5$) shows that this shift is larger when the gluon distribution is harder.

In all figures, except Fig. 11, we have presented the results only for $n_g = -0.5$. However, for $n_g = 0$ there is very little change in our results - at least in the range of variables where data are available.

Finally, we give in Fig. 12 the fractions of the Pomeron momentum carried by quarks and gluons. With $n_g = -0.5$ ($n_g = 0$). For $Q_0^2 = 5$ GeV$^2$, we have

$$N_g = \int_0^1 \beta\, g^P(\beta) d\beta = 0.27 \quad (0.15) \tag{9}$$

$$N_q = \int_0^1 \beta \sum_i \left( q_i^P(\beta) + \bar{q}_i^P(\beta) \right) d\beta = 0.13 \quad (0.13) \quad . \tag{10}$$

We see that for both values of $n_g$ the momentum fraction carried by gluons is larger than the one carried by quarks, indicating that the Pomeron has a substantial gluonic component with a very hard gluon distribution. The energy-momentum sum rule is strongly violated.

## 5. Comparison with other QCD analysis and conclusions

Very recently, QCD-based analysis of the Pomeron structure function and its $Q^2$ evolution have been carried out [13, 22, 23, 24]. In these papers the initial condition in DGLAP evolution equation and, in particular, the gluon momentum distribution is different from ours. However, in all cases conclusions similar to ours are reached, namely, the $Q^2$ dependence of the data can be reproduced only if the gluon momentum distribution in the Pomeron is hard - the gluons carrying more momentum than the quarks. Note that in Refs. [22, 23] the gluon normalization relative to that of the quarks is determined from the momentum sum rule. This leads to a large normalization constant for gluons. It is known from UA8 results [18] that the normalization of partons satisfying the momentum sum rule leads to absolute predictions which exceed experiment by a factor close to five (see also Ref. [9]).

In conclusion, recent HERA data confirm the predictions of the CKMT model of hard diffraction based on Regge theory, factorization, and the assumption that not only the Pomeron intercept but also the triple Reggeon couplings are the same in soft and hard diffraction. Although a QCD analysis of the $Q^2$-dependence of the present data does not allow to extract the shape of the gluon distribution in the Pomeron, it requires a hard gluon distribution - the gluons carrying substantially more momentum than the quarks. This conclusion can be reached without reference to the momentum sum rule, using a normalization constant of the gluon distribution determined from Regge factorization.

**Acknowledgements**

It is a pleasure to thank P. Aurenche and M. Fontannaz for providing an updated versions of the code for DGLAP evolution used in Refs. [19, 20], and R. Engel for providing the version of the same code used in Ref. [14], adapted to our initial condition.

The present work has been realized with the help of an INTAS contract 93-0079. One of the authors (C. M.) has benefitted of a EEC postdoctoral project ERBCHBICT 930547.

# References

[1] M. Derrick et al (Zeus Collaboration), Phys. Lett. **B315**, 481 (1993) ; **B332**, 228 (1994) ; **B338**, 477 (1994).

[2] B. Foster (Zeus Collaboration), Workshop on DIS and QCD, Paris, April 1995.

[3] T. Ahmed et al (H1 Collaboration), Nucl. Phys. **B429**, 477 (1994).

[4] T. Ahmed et al (H1 Collaboration), Phys. Lett. **B348**, 681 (1995) .

[5] A. Capella, A. Kaidalov, C. Merino and J. Tran Thanh Van, Phys. Lett. **B343**, 403 (1995).

[6] A. Capella, A. Kaidalov, C. Merino and J. Tran Thanh Van, Phys. Lett. **B337**, 358 (1994).

[7] A. Capella, J. Kaplan and J. Tran Thanh Van, Nucl. Phys. **B97**, 493 (1975).
K. A. Ter-Martirosyan, Sov. J. Nucl. Phys. **44**, 817 (1986).
A. B. Kaidalov, K. A. Ter-Martirosyan and Yu. M. Shabelski, Sov. J. Nucl. Phys. **44**, 822 (1986).

[8] A. Donnachie and P. V. Landshoff, Nucl. Phys. **B303**, 634 (1988).

[9] K. Goulianos, preprint RU 95/E-06 (HEP-PH 950 2356).

[10] G. Ingelman and P. E. Schlein, Phys. Lett. **B152**, 256 (1985).

[11] G. Ingelman and K. Prytz, Z. Phys. **C58**, 289 (1993).

[12] E. L. Berger et al, Nucl. Phys. **B286**, 704 (1987).

[13] N. N. Nikolaev and B. G. Zakharov, Z. Phys. **C53**, 331 (1992).
M. Genovese, N. N. Nikolaev and B. G. Zakharov, preprint KFA-IKP (Th) 1994-37.

[14] R. Engel, J. Ranft and S. Roesler, to appear in Phys. Rev. D.

[15] A. Capella, J. Kaplan and J. Tran Thanh Van, Nucl. Phys. **B105**, 333 (1976).
A. B. Kaidalov, L.A. Ponomarev and K. A. Ter-Martirosyan, Sov. Journal of Nucl. Phys. **44**, 468 (1986).

[16] A. B. Kaidalov, Phys. Rep. **50** (1979) 157.

[17] For a review see E. Reya, Phys. Rep. **B69**, 195 (1981).

[18] A. Brandt et al (UA8 Collaboration), Phys. Lett. **B297**, 417 (1992) and to be published.

[19] A. Devoto, D. W. Duke and J. F. Owens, Phys. Rev. **D27**, 508 (1983).

[20] P. Aurenche, R. Baier, M. Fontannaz, M. N. Kienzle-Focacci and M. Werlen, Phys. Lett. **B233**, 517 (1989), in preparation.

[21] A. Kaidalov, Workshop on DIS and QCD, Paris, April 1995.

[22] T. Gehrmann and W. J. Stirling, DTP/95/26.

[23] K. Golec-Biernat and J. Kwiecinski, Krakow INP Report No 1670/PH.

[24] J. Dainton (H1 Collaboration), Workshop on DIS and QCD, Paris, April 1995.

# Figure Captions

**Fig. 1** Pomeron exchange diagram for single diffraction dissociation of a virtual photon.

**Fig. 2** Pomeron exchange diagram for a double diffraction dissociation in which both the hard current and the target proton are excited.

**Fig. 3** Regge exchange diagram for the proton structure function.

**Fig. 4** Regge exchange diagram for the Pomeron structure function.

**Fig. 5** Initial gluon distribution ($Q_O^2 = 5$ GeV$^2$) and its perturbative QCD evolution for $n_g = -0.5$.

**Fig. 6** Same as Fig. 5 for a $u$ quark.

**Fig. 7** Same as Fig. 5 for an $s$-quark.

**Fig. 8** Same as Fig. 5 for a $c$-quark. The charm distribution is taken to be zero at $Q_0^2$ and is generated, at larger $Q^2$, via DGLAP evolution.

**Fig. 9** The 3-dimensional proton structure function for diffractive events $F_2^D(x, x_P, Q^2)$ is plotted versus $x_P$ at various values of $Q^2$ and $\beta = x/x_P$ and compared with H1 data [4]. The model results are for $n_g = -0.5$. Very similar results are obtained with $n_g = 0$.

**Fig. 10** The 2-dimensional proton structure function for diffractive events $F_2^D(\beta, Q^2)$ is plotted versus $Q^2$ at fixed $\beta$ and versus $\beta$ at fixed $Q^2$ and compared with H1 data [4]. The model results are for $n_g = -0.5$. Very similar results are obtained with $n_g = 0$. In order to take into account the presence of double diffraction in the data, the theoretical result has been multiplied by 1.3.

**Fig. 11** The Pomeron structure function $F_2^P(\beta, Q^2)$, defined in Eq. (4), is plotted versus $\beta$ at several values of $Q^2$ for $n_g = -0.5$ (Fig. 11a) and $n_g = 0$ (Fig. 11b).

**Fig. 12** The functions of the Pomeron momentum carried by quarks ($N_q$) and gluons ($N_g$) defined by eqs. (9) and (10) are plotted versus $Q^2$ for two different values of $n_g$.

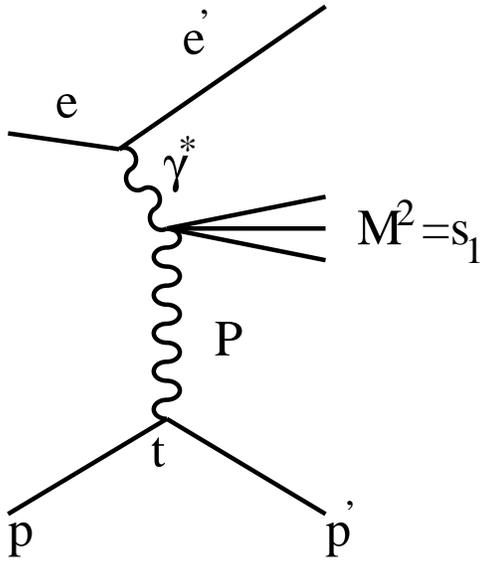

Figure 1

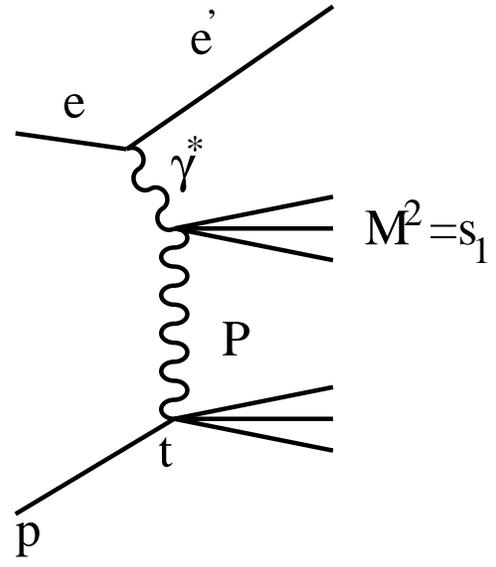

Figure 2

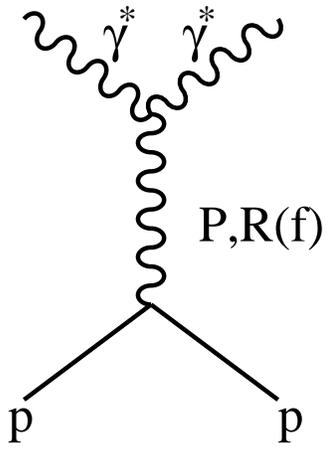

Figure 3

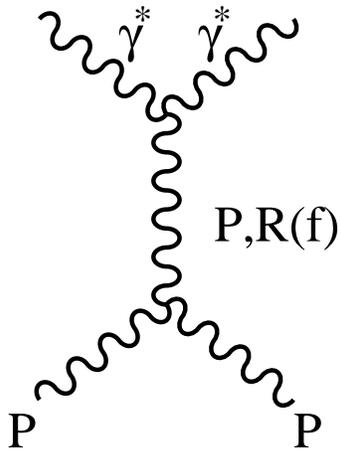

Figure 4

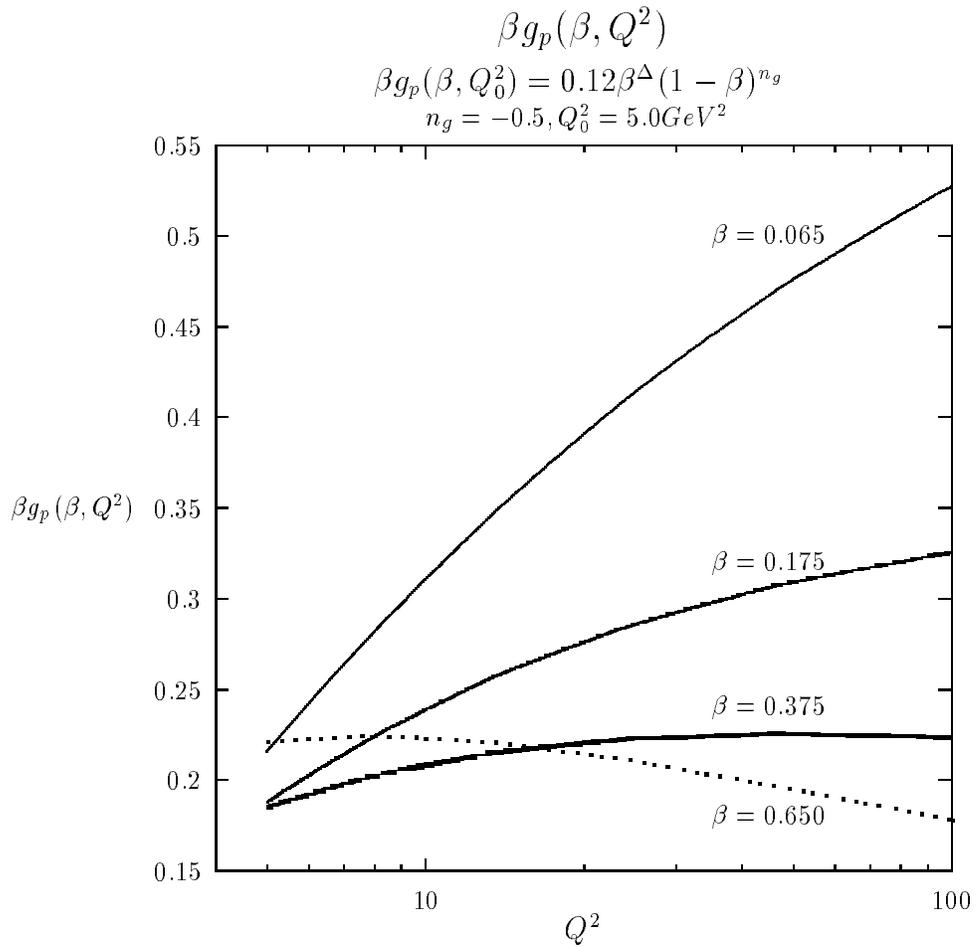

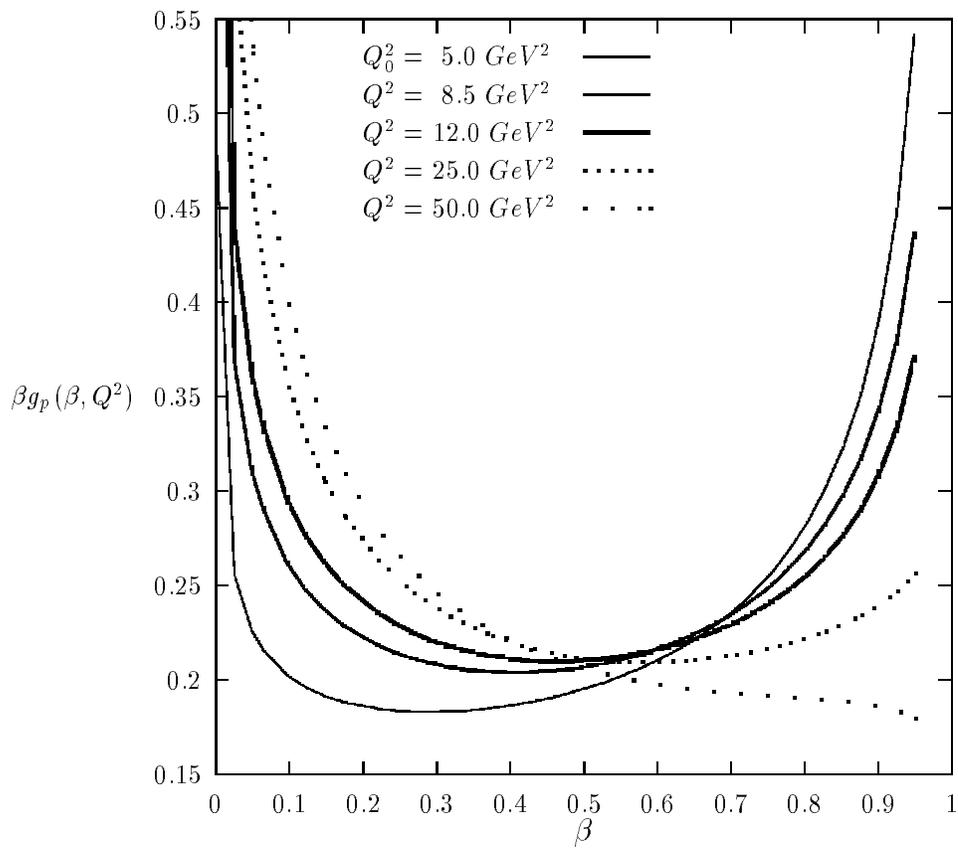

Fig.5

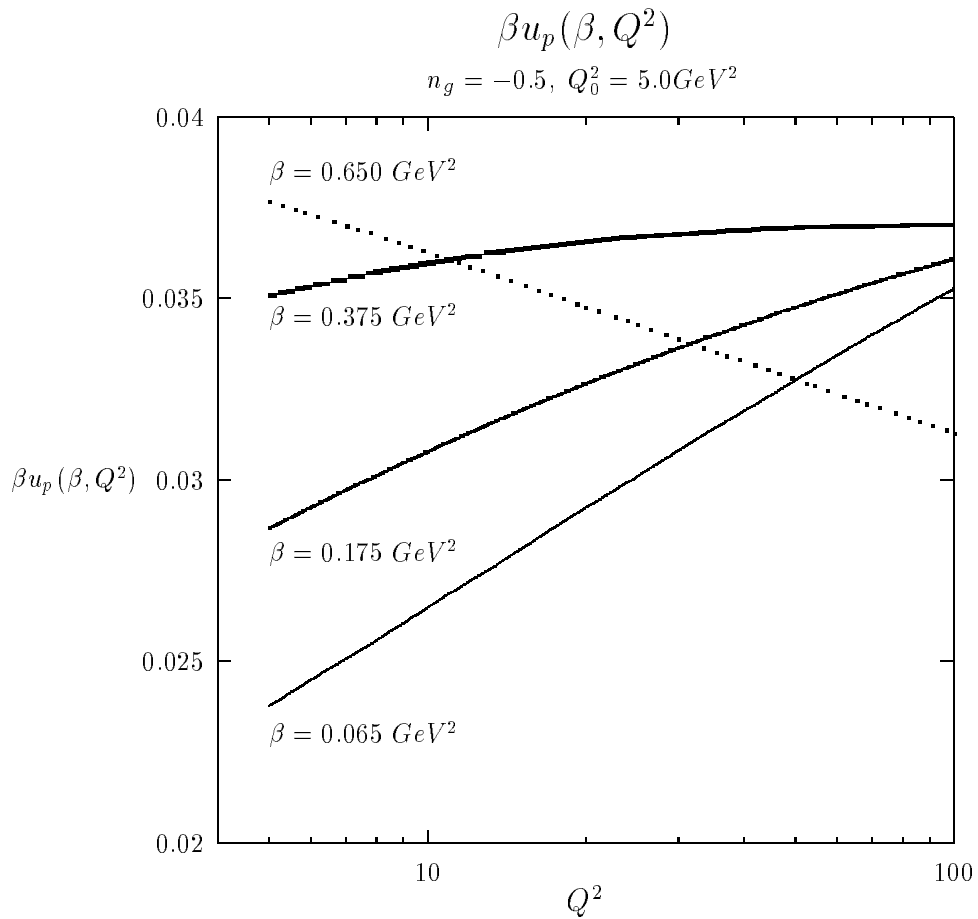

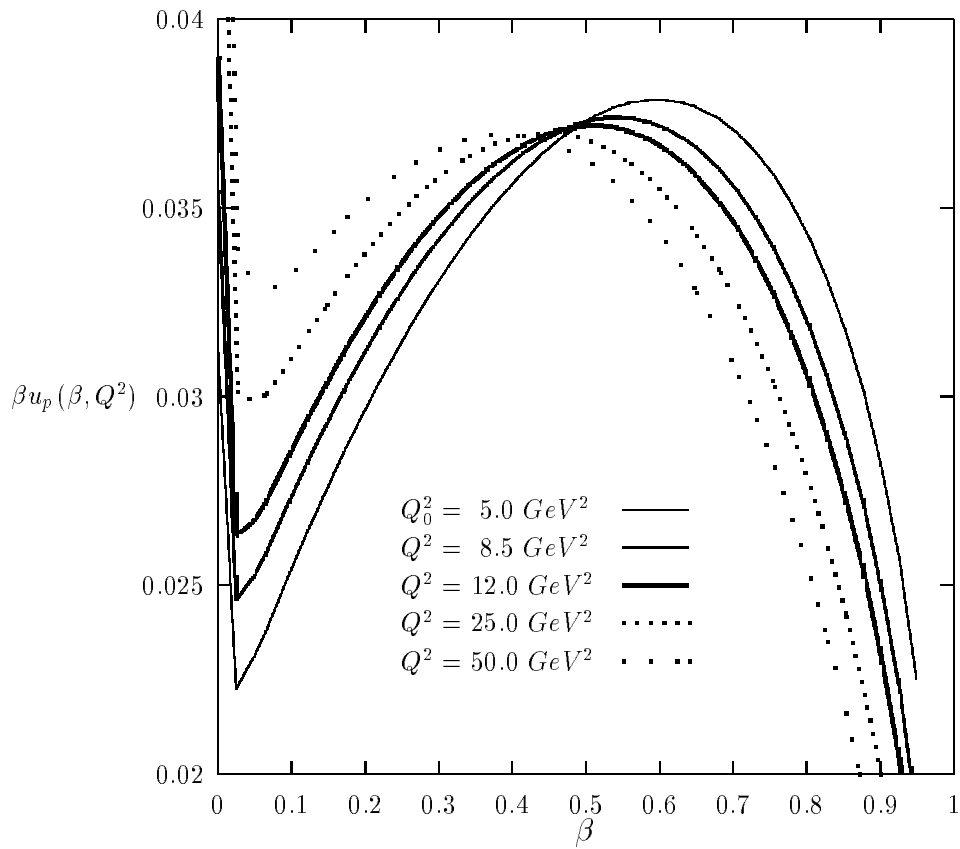

Fig.6

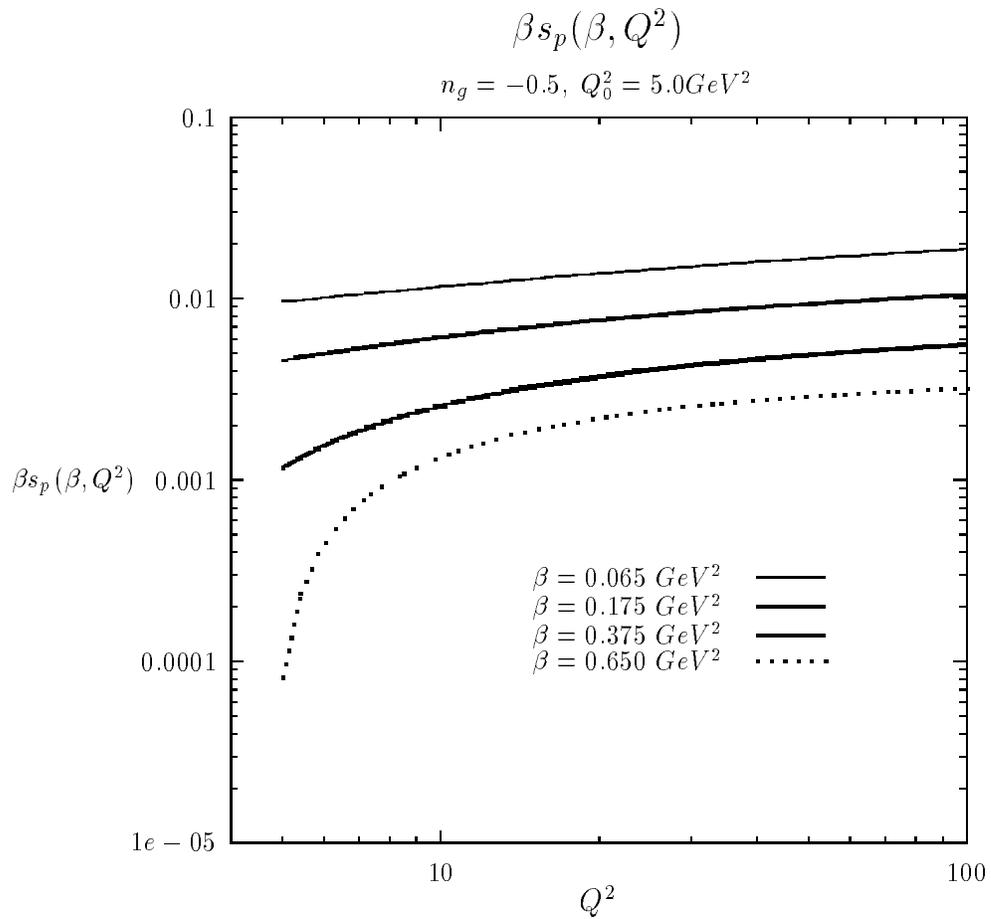

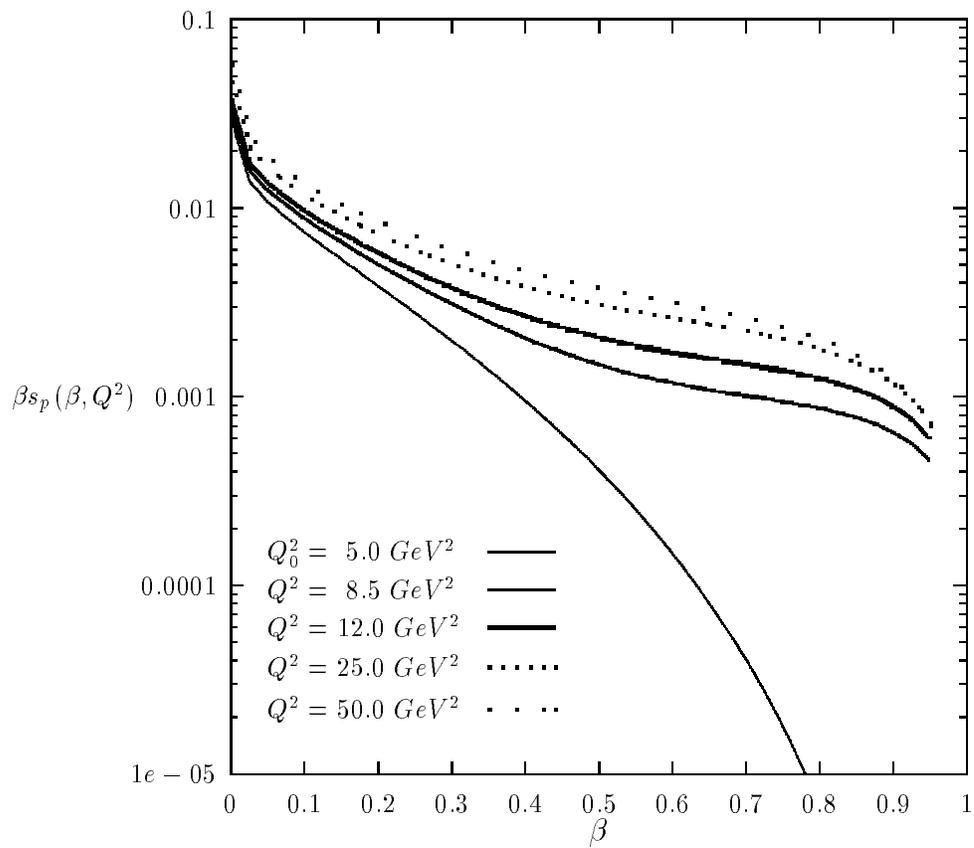

Fig.7

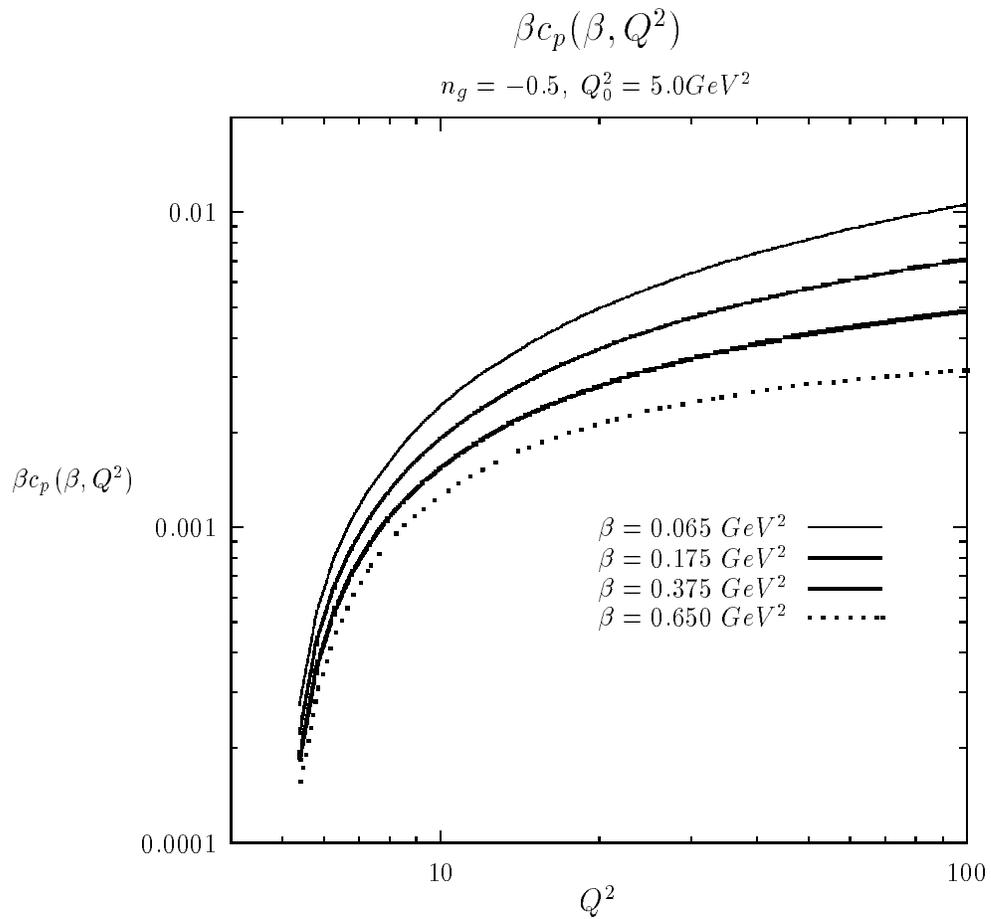

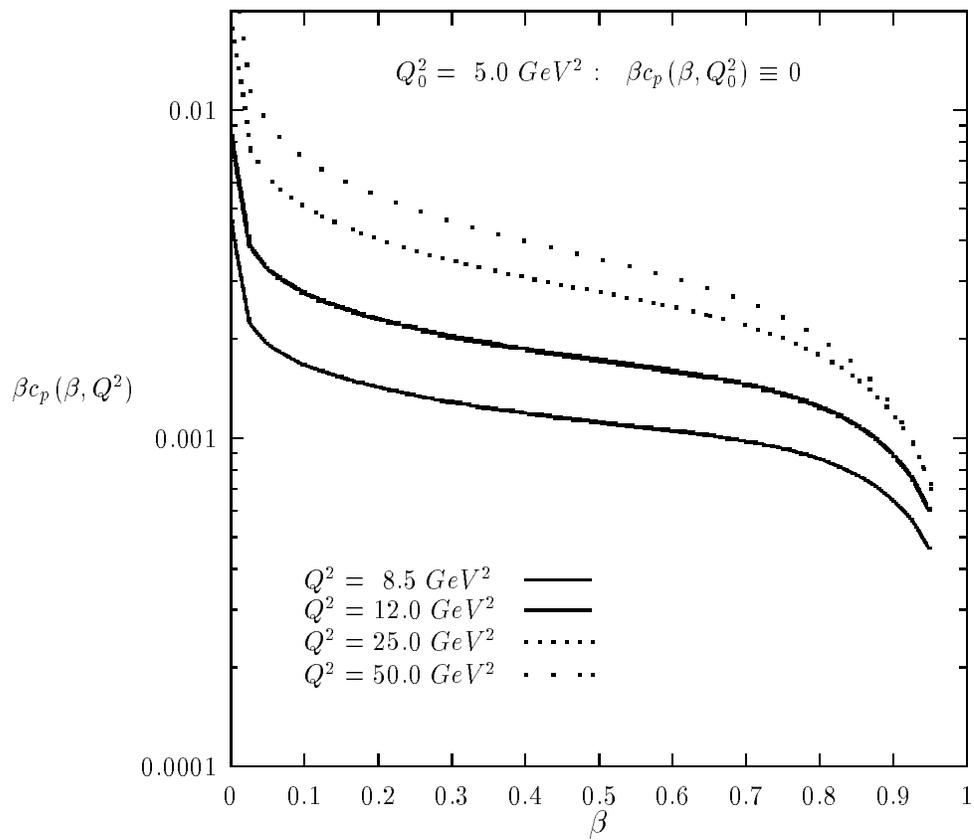

Fig.8

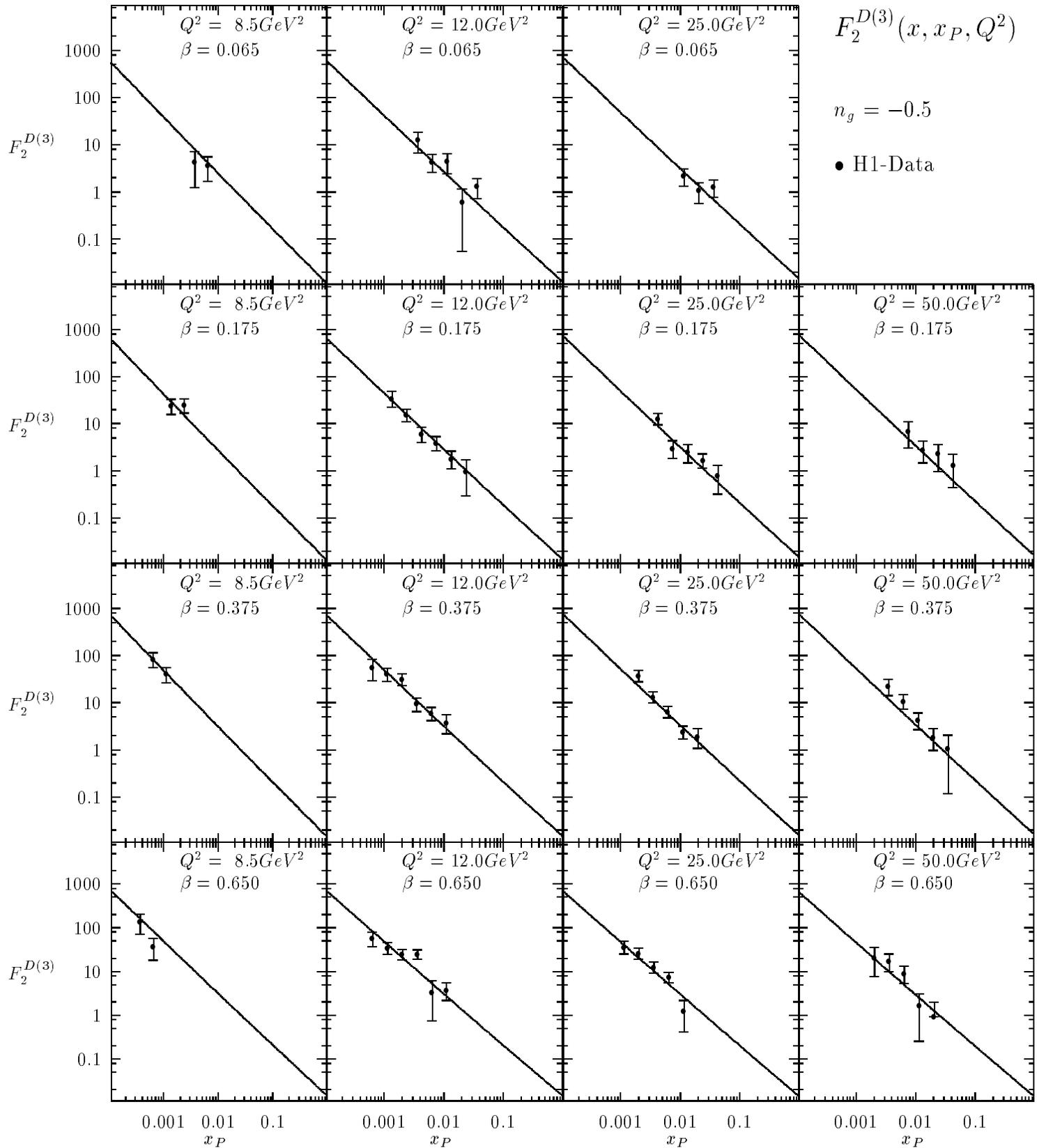

Fig.9

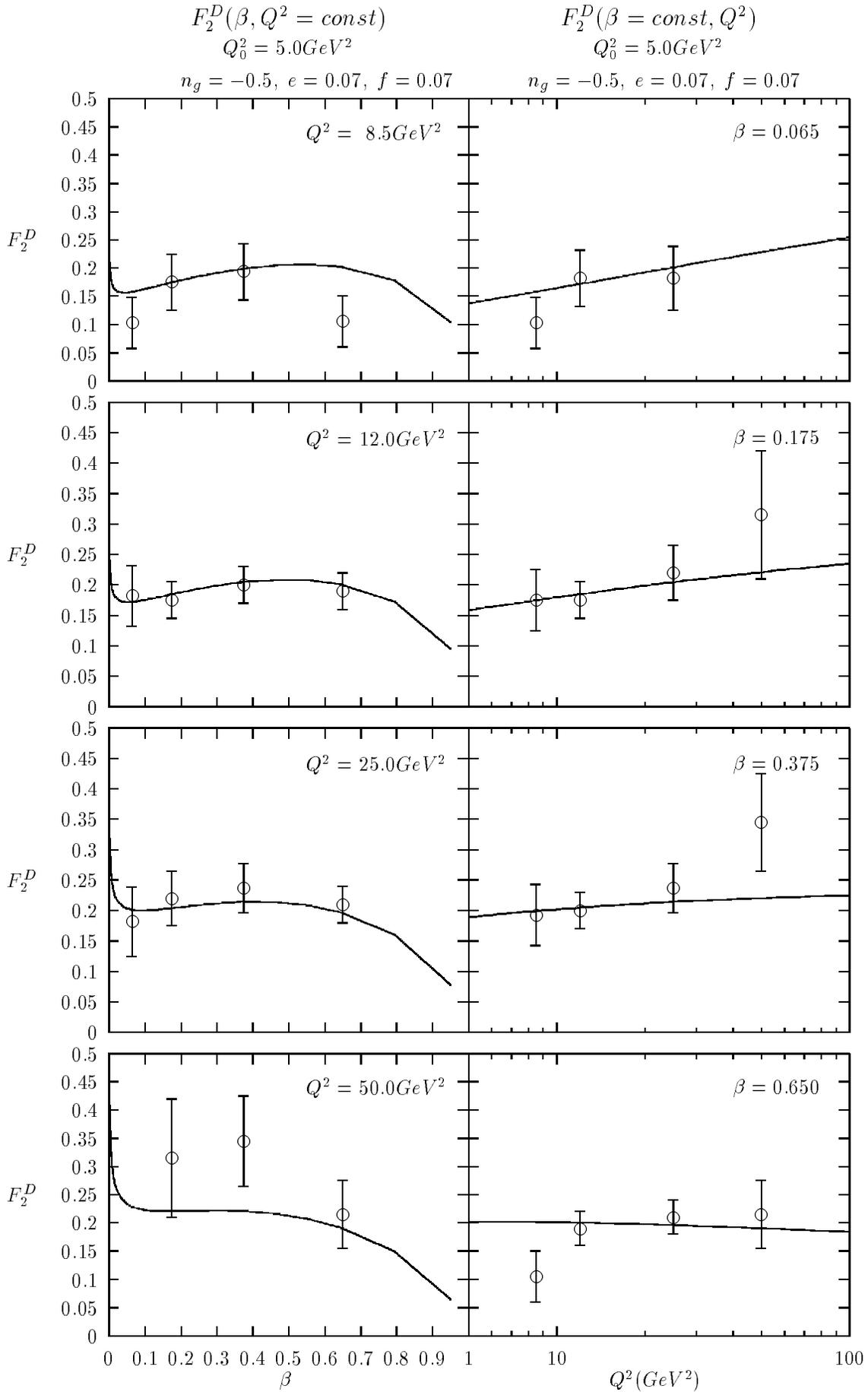

Fig.10

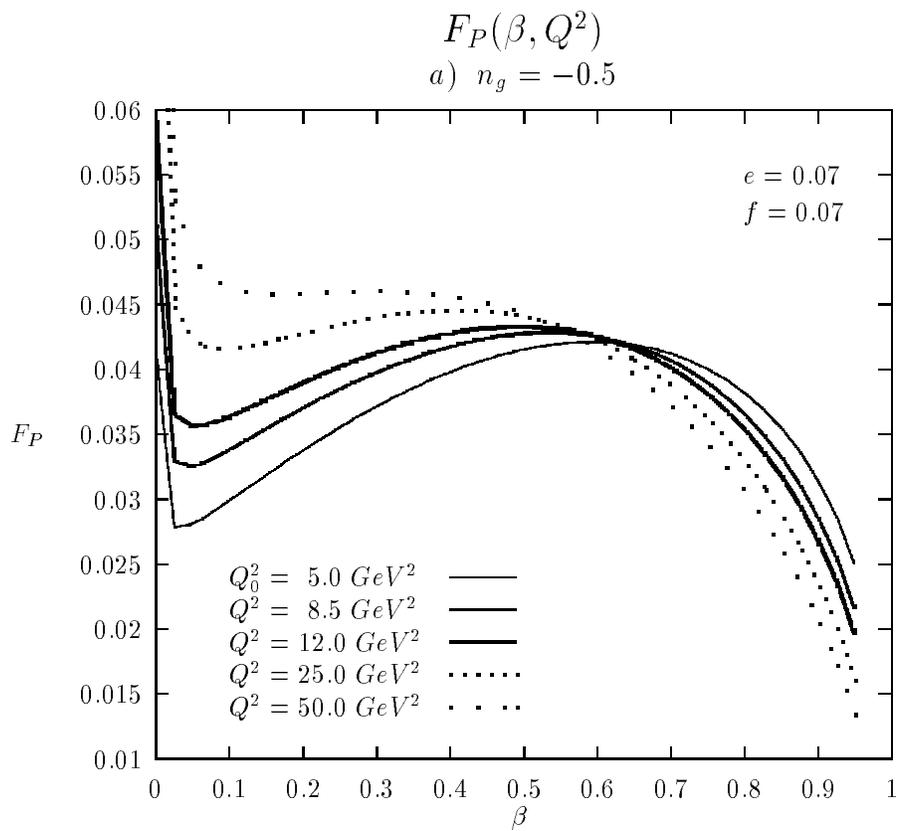
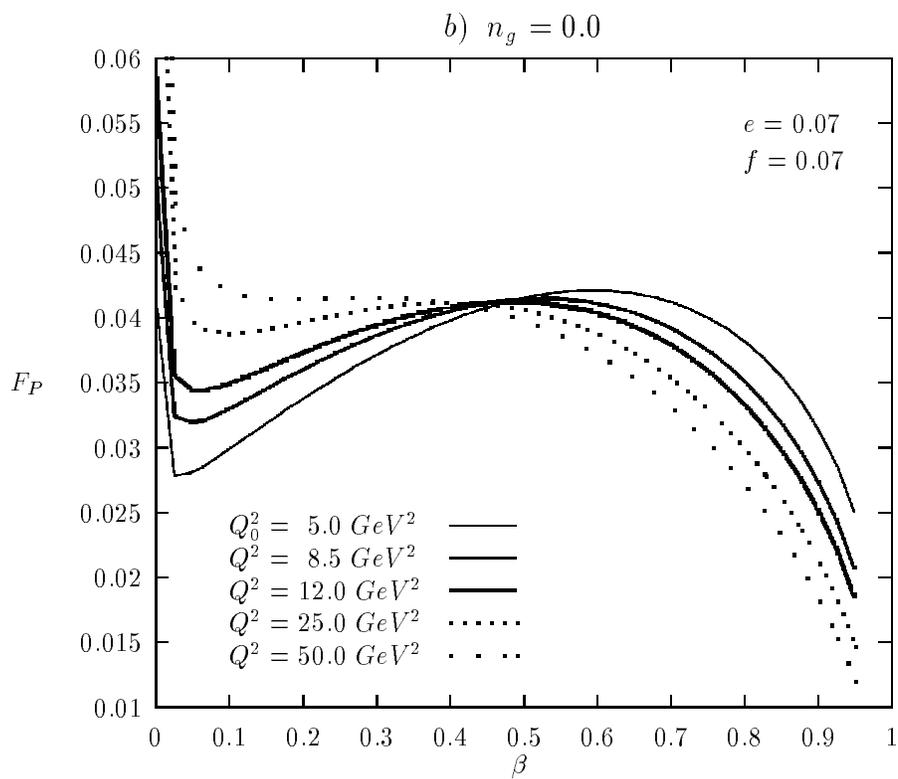

Fig.11

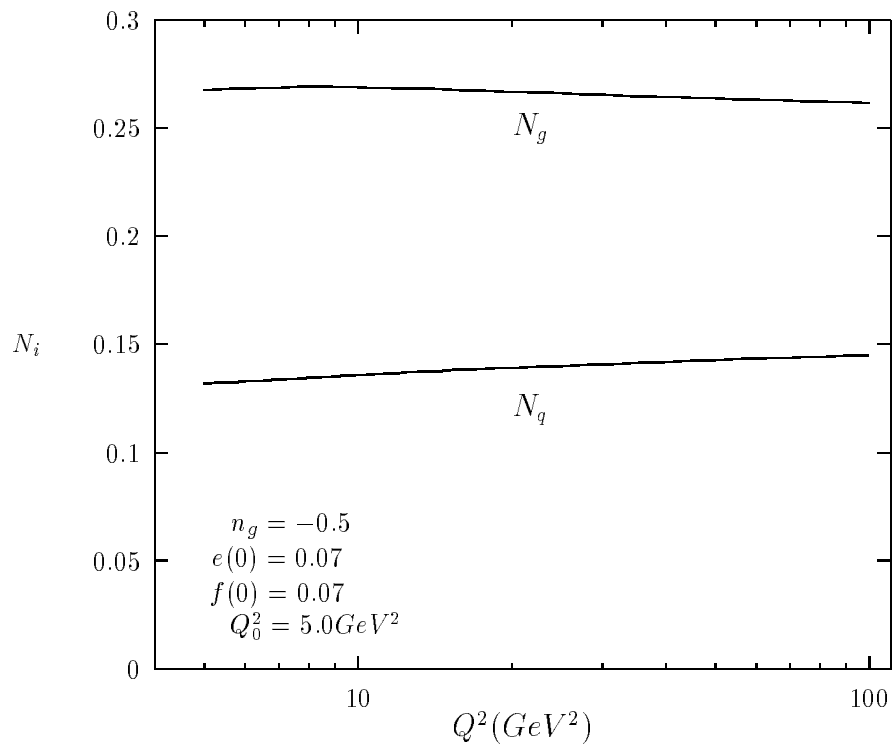

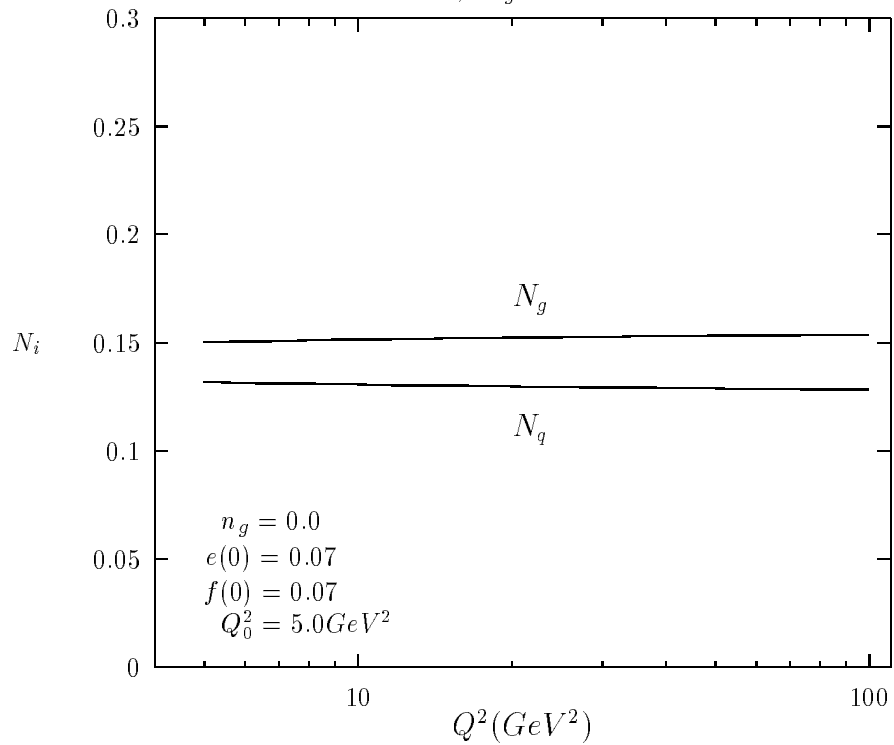

Fig.12